*Quantum processor-inspired machine learning in the biomedical sciences*

Richard Y. Li[1,2,3], Sharvari Gujja[4,5], Sweta R. Bajaj[4,5], Omar E. Gamel[4], Nicholas Cilfone[4], Jeffrey R. Gulcher[6], Daniel A. Lidar[§1,3,7,8], Thomas W. Chittenden[§4,5,9]

[1]Department of Chemistry, University of Southern California, Los Angeles, CA, USA

[2]Computational Biology and Bioinformatics Program, Department of Biological Sciences, University of Southern California, Los Angeles, CA, USA

[3]Center for Quantum Information Science & Technology, University of Southern California, Los Angeles, CA, USA

[4]Computational Statistics and Bioinformatics Group, Advanced Artificial Intelligence Research Laboratory, Genuity Science, Boston, MA, USA

[5]Complex Biological Systems Alliance, Medford, MA, USA

[6]Cancer Genetics Group, Genuity Science, Boston, MA, USA

[7]Department of Electrical Engineering, University of Southern California, Los Angeles, CA, USA

[8]Department of Physics and Astronomy, University of Southern California, Los Angeles, CA, USA

[9]Division of Genetics and Genomics, Boston Children's Hospital, Harvard Medical School, Boston, MA, USA

[§] Corresponding authors

**Materials and Correspondence**
Thomas W. Chittenden, PhD, DPhil, PStat
Genuity Science
90 Canal Street, Suite 120
Boston, MA 02114
Phone: (617) 250 3500
Email: tom.chittenden@genuitysci.com

Daniel A. Lidar, PhD
University of Southern California
920 Bloom Walk
Los Angeles, CA 90089
Phone: (213) 740 0198
Email: lidar@usc.edu

**Summary**

Recent advances in high-throughput genomic technologies coupled with exponential increases in computer processing and memory have allowed us to interrogate the complex aberrant molecular underpinnings of human disease from a genome-wide perspective. While the deluge of genomic information is expected to increase, a bottleneck in conventional high-performance computing is rapidly approaching. Inspired in part by recent advances in physical quantum processors, we evaluated several unconventional machine learning (ML) strategies on actual human tumor data; namely "Ising-type" methods whose objective function is formulated identically to simulated annealing and quantum annealing, as well as a restricted Boltzmann machine. Here we show for the first time the efficacy of multiple Ising-type ML algorithms for classification of high-dimensional, multi-omics human cancer data from the Cancer Genome Atlas. To assess algorithm performance, we compared these classifiers to a variety of standard ML methods. Our results indicate the feasibility of using Ising-type ML to provide competitive classification of human cancer types and associated molecular subtypes and superior performance with smaller training datasets, thus providing compelling empirical evidence for the potential future application of unconventional computing approaches in the biomedical sciences.

**Introduction**

With the rapid expansion of high-throughput genomic technologies there exists a multitude of 'omics' data, which allows researchers to now investigate the casual molecular drivers of complex human disease with a systems biology approach. Over the past two decades, numerous studies have shown the utility of statistical ML strategies to classify human malignancies, hypothesize unknown clinical subtypes, and make prognostic predictions based on omics datasets (Golub et al., 1999; Nevins and Potti, 2007). Moreover, integrated 'multi-omics' approaches have proved effective in deriving meaningful biological insights into the etiological and prognostic complexity of human cancers (Hoadley et al., 2018; Hoadley et al., 2014; Robinson et al., 2017; Uhlen et al., 2017). While these studies highlight the potential of omics-based analytics to drive innovative new therapies based on unique molecular signatures, several well-documented issues, including correlation bias, feature dependency, and multicollinearity still hamper statistical optimization for the analysis and robust classification of high-dimensional complex biological datasets (Toloşi and Lengauer, 2011).

To address some of these statistical computing limitations, we present a class of unconventional "Ising-type" machine learning algorithms, inspired by quantum computing. As a rapidly emerging technology, quantum computing promises to enhance performance of certain classes of statistical computing and ML tasks, such as classification, regression, generation, and resampling. In this nascent discipline, proposals for several quantum ML algorithms have been developed, including quantum principal component analysis (Lloyd et al., 2014) and quantum support vector machines (Rebentrost et al., 2014) and Boltzmann machines (Biamonte et al., 2017). These proposals have generated interest in the scientific community and in the general public for their potential to address computationally intractable tasks and to model more complicated data distributions. One of the unconventional machine learning approaches used in this study, quantum annealing with processors made by D-Wave (Harris et al., 2010; Johnson et al., 2011; Lanting et al., 2014), feature more than 2000 qubits are becoming large enough to solve real-world problems (Rieffel et al., 2015), perform quantum simulation (King et al., 2018), and compete with classical optimization algorithms (Albash and Lidar, 2018). While the computational role of quantum effects in these processors remains controversial and the subject of intensive study, quantum annealing is currently one of the few paradigms of quantum computing that are approaching a scale useful for practical applications.

Using high-dimensional, multi-omics human cancer data from the Cancer Genome Atlas (TCGA), we framed a classification problem in such a way that is amenable to solving with Ising-type approaches. The Ising-type methods must be formulated as a quadratic unconstrained optimization (QUBO) or equivalently, an Ising Hamiltonian $H(w) = w^\mathsf{T}h + w^\mathsf{T}Jw$, where $\boldsymbol{w}$ is a vector of weights, and $h$ and $J$ represent a vector and matrix, respectively. We compared Ising-type approaches to standard machine learning approaches for both binomial and multiclass experimental designs. Although previous studies have applied quantum annealing and other Ising models to model protein folding (Perdomo-Ortiz et al., 2012), transcription factor DNA binding (Li et al., 2018), and classification of lung cancer data with microarray data (Jain et al., 2020), our analysis is the first of integrated, genome-wide multi-omics human cancer data. In the course of our study we found that the Ising models all perform similarly. Our results show that in most cases when using relatively large amounts of high-dimensional multi-omics training data, the Ising-type methods are comparable to standard supervised ML approaches. However, for smaller training datasets of equivalent dimensionality, Ising models statistically outperform established classification strategies. We also assessed the weights returned by the Ising models and found reasonable interpretability and generalizability of biological information. Overall our results demonstrate the current utility and limitations of Ising models applied to the analysis of high-dimensional omics data, and point to a general class of algorithms that may be useful when training data is limited.

**Results**

We assessed the performance of annealing-based Ising ML algorithms on several TCGA datasets to identify comparative advantages for the Ising approaches. In this machine learning survey, we compared performance of Ising models to the following commonly used ML algorithms: Least Absolute Shrinkage and Selection Operator (LASSO) (Tibshirani, 1996), Ridge regression (Ridge) (Hoerl et al., 1975; Hoerl and Kennard, 1970), Random Forest (RF) (Breiman, 2001; Breiman et al., 1993), Naïve Bayes (NB) (Hastie et al., 2016; Ng, 2008), and a Support Vector Machine (SVM) (Boser et al., 1992; Cortes and Vapnik, 1995). TCGA data, including exome DNA variation, RNA-seq, DNA methylation, miRNA, and CNVs, were retrieved, pre-processed, and normalized, resulting in an average of 70,504 gene features for five binomial and one multiclass six-cancer TCGA dataset comparisons. We performed dimensionality reduction with principal component analysis (see Supplemental Methods), retaining the top 44 principal components for the binomial datasets and 13 principal components for the six-cancer dataset. The number of principal components was chosen based on the largest number of features that could be accommodated on existing quantum annealing hardware. An overview of our data analysis strategy is presented in Figure 1.

Quantum annealing (QA) was implemented on D-Wave physical quantum processors (see Supplemental Methods). As mentioned, D-Wave only admits problems formulated as a quadratic unconstrained binary optimization (QUBO) problem, or equivalently, an Ising Hamiltonian, generically written as $H(w) = w^\mathsf{T}h + w^\mathsf{T}Jw$, where $\boldsymbol{w}$ is a vector of weights, $J$ is the matrix of interactions, and $\boldsymbol{h}$ is the local fields. The goal of the learning procedure is to find an optimal set of weights that minimizes the energy of the Ising Hamiltonian; i.e., find $w^* = argmin_w H(w)$. The global optimum of the Ising problem is in general difficult to determine (Barahona, 1982). Classification by using quantum annealing to solve Ising problems has been formulated before (Pudenz and Lidar, 2013). In the present work, we developed a novel approach that can be used to solve classification problems directly. Our strategy stems from multinomial regression, which reduces to logistic regression when there are two classes (see Supplemental Methods for the mapping to an Ising problem). We compared performance to several other Ising models that use the same objective function as D-Wave; i.e., problems formulated as an Ising Hamiltonian: simulated annealing (SA), Random, and Field. Simulated annealing (Kirkpatrick et al., 1983) is a well-known heuristic optimization algorithm that uses Metropolis updates and a (fictitious) temperature schedule to optimize a target objective function. For Random, we randomly generated candidate weights, sorted them by their Ising energy and selected the best performing weights. For Field, we disregarded $J$, the coupling terms, and performed an optimization only over $h$, the local fields (See Supplemental Methods for more details of all classical, quantum and quantum-inspired algorithms). Both Random and Field were introduced and used as simple benchmarks against which we tested the SA and QA approaches.

Lastly, we compared performance of the annealing-based models to a restricted Boltzmann machine (RBM), which is also based on an Ising model. Note however, that the annealing-based Ising models described above are purely supervised learning approaches; i.e., formulated explicitly with a response variable to be predicted (in this case, the class of cancer). Boltzmann machines generally seek to explicitly model a data distribution over the inputs as a Boltzmann distribution by incrementally adjusting the $h$'s and $J$'s; whereas in our formulation the $h$'s and $J$'s are fixed given the input training data, and the mechanism of learning is to obtain solutions that minimize the Ising Hamiltonian, rather than to accurately reproduce a distribution. Accordingly, we will use "all Ising models" to refer to both the annealing-based Ising approaches and the RBMs, and "annealing-based" Ising approaches to refer exclusively to the annealing-based approaches.

### *Binomial and Multinomial Classification*
In this section, we present classification results for five binomial TCGA cancer dataset comparisons: kidney renal clear cell carcinoma (KIRC) vs. kidney renal papillary cell carcinoma (KIRP); lung adenocarcinoma (LUAD) vs. lung squamous cell carcinoma (LUSC); breast invasive carcinoma (BRCA) vs. matched normal breast tissue (normal); estrogen receptor positive (ERpos) vs. estrogen receptor negative (ERneg) breast cancers; and luminal A (LumA) vs. luminal B (LumB) breast cancers. We also present findings relative to a six-cancer multiclass classification strategy for human brain, breast, kidney, lung, liver, and colorectal cancer types (see Supplemental Spreadsheet 1-S1 for the sample sizes of each dataset). We assessed the relative classification performance of the five standard ML models (LASSO, Ridge, RF, NB, and SVM), one quantum algorithm (D-Wave), three annealing-based Ising algorithms (simulated annealing (SA), Random, and Field), and one Ising model (RBMs) for all binomial and multiclass TCGA comparisons.

Figure 2 presents comparisons of all ten classifiers for the five binomial datasets. We used four statistical metrics to assess classification performance: accuracy, balanced accuracy, ROC area under the curve (AUC), and F1 score. The four metrics were independently averaged over 100 unique training and test sets for each classifier (see Supplemental Methods). Mean ± SEM for each metric are presented on the y-axis of each figure inset. See Table 1 for the values of the balanced accuracy and Supplemental Tables 1-3 for accuracy, AUC, and F1 score. Relative classification performance was determined by mean balanced accuracy and presented in ranked order on the x-axis of each figure inset. Nonparametric Wilcoxon signed-rank tests were used to assess statistical significance among the ten classifiers relative to the four performance metrics. Bonferroni correction was used to adjust for multiple testing error. For each comparison, we found a standard ML approach outperformed both quantum and classical annealing across all four metrics of performance. However, for several comparisons, at least one of the Ising-type algorithms performed nearly as well as the best classical method. For example, while RF statistically outperformed ($0.99 \pm 0.002$) all other methods for the BRCA vs. Normal comparison, Random, SVM, RBM, SA, and LASSO showed no statistical differences in performance ($0.98 \pm 0.002$; $0.98 \pm 0.002$; $0.98 \pm 0.002$; $0.98 \pm 0.003$; $0.98 \pm 0.002$). Similarly, for the LumA vs. LumB comparison, we found LASSO performed best ($0.76 \pm 0.006$); however, Random, D-Wave, SA, Ridge, and Field are nearly identical in terms of balanced accuracy ($0.75 \pm 0.006$; $0.75 \pm 0.006$; $0.75 \pm 0.006$; $0.74 \pm 0.006$; $0.74 \pm 0.006$). For the three other comparisons (ERpos vs. ERneg; KIRC vs. KIRP; LUAD vs. LUSC), the annealing-based Ising algorithms statistically underperformed the best standard ML algorithm in each comparison, though RBM performed well. While D-Wave performed similarly to RF and NB in the KIRC vs. KIRP comparison ($0.94 \pm 0.002$ vs. $0.94 \pm 0.002$; $0.94 \pm 0.002$; corrected p-value = 1), it was statistically inferior to SVM ($0.94 \pm 0.002$ vs. $0.98 \pm 0.001$; corrected p-value = $5.96 \times 10^{-24}$). Overall Field was one of the poorest performing methods relative to the four metrics assessed; however, it performed relatively well on the LumA vs. LumB dataset ($0.74 \pm 0.006$). The quantum and classical Ising-type classification results indicate the utility of framing an overall classification strategy as an Ising problem.

Though the Ising-type algorithms generally underperformed the standard ML methods assessed for these comparisons, the Ising-type classifiers performed well on the LumA vs. LumB comparison. Moreover, as with all the standard ML methods used in this work, the most informative feature for classification predicted by the annealing-based Ising models is the first principal component, indicating the Ising models also assigned the greatest weight to the features that account for the most variation within the data. This is consistent with previous results where D-Wave was able to extract a motif for protein-DNA binding that agreed with classical results (Li et al., 2018).

Finally, to determine the utility of Ising-type methods on a larger, multiclass classification experimental design, we evaluated classification performance of the standard and Ising-type ML algorithms on a six-cancer, multiclass TCGA dataset. The six TCGA cancer types included brain, breast, kidney, lung, liver, and colorectal cancers (see Supplemental Spreadsheet 1-S1 for the sample size of this six-cancer dataset). With the exception of multiclass AUC ($0.99 \pm 0.0$), performance metrics for standard ML approaches were superior to all Ising models for this larger, multiclass dataset (see also Supplemental Figure S1). We, therefore, focused our efforts on further evaluating the efficacy of all the Ising models on the five binomial comparisons described above.

### *Performance Dependence on Training Set Size*

Based on previous work indicating quantum and classical Ising-type approaches are superior to standard ML classifiers on small training set sizes (Li et al., 2018; Mott et al., 2017; Willsch et al., 2019), we systematically reduced the training set data for the LumA vs. LumB human breast cancer comparison into 16 separate partition sizes to evaluate classifier performance (see Supplemental Methods). We first divided the entire LumA vs. LumB breast cancer dataset (311 breast tumor samples) into a training set representing 80% of the initial dataset (250 breast tumor samples) and a testing set equal to 20% of the initial dataset (61 breast tumor samples). From this, we randomly selected incrementally smaller, class-balanced data partitions from 95% to 20% of the original training set data. Due to the complexity and computational expense of this experimental design, we trained each of the ten classifiers described above over only 50 unique training sets, randomly drawn from the 250 breast tumor samples of the initial training data, for each individual training set partition. We then validated performance of each classifier on the original, held-out test set of 61 breast tumor samples. As above, nonparametric Wilcoxon signed-rank tests were used to assess statistical significance among the ten classifiers relative to the four performance metrics, and Bonferroni correction was used to adjust for multiple testing error. Results in Figure 3 are presented as mean ± SEM for averaged balanced accuracies across the entire training set size spectrum.

At 25-40% of the original training data (63 to 100 breast tumor samples), the mean balanced accuracies of the five Ising models (D-Wave, SA, Random, Field, and RBM) were statistically superior to the mean balanced accuracies of the five standard ML algorithms (LASSO, NB, RF, Ridge, SVM). For example, at 25% of the initial training data D-Wave statistically outperformed SVM, the top standard ML method ($0.74 \pm 0.007$ vs. $0.70 \pm 0.007$; corrected p-value = $1.94 \times 10^{-3}$), as did the other Ising-type methods. Classification performance for all standard ML methods (SVM, LASSO, NB, RF, Ridge) steadily decreased after a reduction to 50% of the original training data (125 breast tumor samples), whereas we found significantly less reduction in mean balanced accuracies for the five quantum and classical Ising models across the entire training set size spectrum. Furthermore, the Ising models showed a relatively minimal reduction in performance at 95% vs. 20% of original training data ($0.76 \pm 0.004$ vs. $0.73 \pm 0.007$; corrected p-value = 0.092 for SA's performance) as compared to LASSO ($0.75 \pm 0.002$ vs. $0.60 \pm 0.01$; corrected p-value = $3.32 \times 10^{-11}$). Moreover, all five standard ML methods associated with a significantly higher degree of overfitting than the Ising model classification approaches, an issue that has also historically plagued the analysis of genomic data. Supplemental Figure S2a indicates significantly less statistical shrinkage relative to test data for the Ising models algorithms across all fractions of training data for the LumA vs. LumB comparison. As an example, 20% of the training data, although RBM and SVM perform fairly similarly in terms of the balanced accuracy on the test set ($0.73 \pm 0.007$ vs. $0.71 \pm 0.007$, p-value = 0.016), the overfitting, as measured as the difference between the training and test balanced accuracy, is significantly higher for SVM than for RBM ($0.29 \pm 0.007$ vs. $0.17 \pm 0.007$, p-value = $9.37 \times 10^{-14}$).

To assess the generality of this finding that Ising-type methods may perform better than standard ML approaches with a small amount of training data, we performed the same analysis on the ERpos vs. ERneg breast cancer and the six-cancer, multiclass datasets. As both datasets were significantly larger than the LumA vs. LumB comparison, we reduced each to a much smaller percentage of initial training set size. Supplemental Figure S3a presents mean balanced accuracies from 95% to 10% of the original training data (730 to 77 breast tumor samples) for the ERpos vs. ERneg comparison. We found the same result in classification performance for all ten classifiers; namely, a decrease in performance for the small training data for the standard ML methods, but little or no change for the Ising models. Unlike the LumA vs. LumB comparison, the Ising models showed no statistical loss in performance from 95% vs. 10% of original training data ($0.84 \pm 0.004$ vs. $0.84 \pm 0.002$; corrected p-value = 1 for SA); whereas RF dropped from $0.86 \pm 0.002$ to $0.81 \pm 0.008$ (corrected p-value = $1.13 \times 10^{-05}$). Similar to the LumA vs. LumB comparison, Supplemental Figure S2b indicates that the Ising models generally have less over-fitting across many of the training fractions; SVM had a higher degree of overfitting compared to SA ($0.14 \pm 0.006$ vs. $0.02 \pm 0.005$, p-value=$8.48 \times 10^{-17}$).

Analysis of the six-cancer, multiclass dataset further confirmed ERpos vs. ERneg findings. While Supplemental Figure S3b shows that the standard ML methods significantly outperformed the Ising-type methods, here again we found no statistical reduction in D-Wave performance ($0.92 \pm 0.001$ vs. $0.91 \pm 0.002$; corrected p-value = 1) from 95% (3035 tumor samples) to 5% (163 tumor samples) of initial training set size, though RBMs did better than the annealing-based Ising approaches. Comparatively, we again found a significant reduction in classification performance for LASSO ($0.992 \pm 0.0001$ vs. $0.978 \pm 0.001$; corrected p-value = $9.89 \times 10^{-16}$) on this multiclass cancer dataset. In addition, SA also exhibited a significant performance drop relative to D-Wave at the low end of training data fraction, although this feature is temperature dependent: by modifying SA's final temperature it can be made to perform as well as D-Wave. This is concordant with previous binomial qML studies (Kirkpatrick et al., 1983; Pudenz and Lidar, 2013).

In summary, all methods (with the exception of NB) converged to roughly the same balanced accuracy at high training data fraction, but at low fraction all Ising models performed better on three distinct datasets. These findings go beyond previous work (Kirkpatrick et al., 1983; Pudenz and Lidar, 2013) and further bolster the case for the utility of framing an overall classification strategy as an Ising problem. Moreover, robust classification of small, high-dimensional omics datasets with Ising models provides a potential new avenue to evaluate patient response in early-phase clinical drug trials or in other genome-wide datasets with relatively small numbers of patients or animal models.

### *Gene-level Classification*
To assess performance of the Ising-type methods at the gene level, we used the 44 most informative genes, by PCA loading of the first principal component (PC1), from the original training set described in the previous sections for the LumA vs. LumB breast cancer dataset. Results are presented in Figure 4a. The four metrics were independently averaged over 100 unique training and test sets for each of the ten classifiers. Nonparametric Wilcoxon signed-rank tests were again used to assess statistical significance for the four metrics relative to the ten classifiers. As above, Bonferroni correction was used to adjust for multiple testing error. Here we found a significant increase in mean balanced accuracies for all ten classifiers at the gene level compared to PCA feature-based classification. For example, RF performed significantly better at the gene level as compared to PC level ($0.83 \pm 0.007$ vs. $0.65 \pm 0.008$; corrected p-value = $3.02 \times 10^{-31}$). We also found that Random ($0.81 \pm 0.005$), SA ($0.80 \pm 0.005$), and D-Wave ($0.80 \pm 0.006$) slightly outperformed three of the five standard ML approaches: SVM ($0.79 \pm 0.005$), NB (0.79 ± 0.006), and LASSO (0.77 ± 0.005). To confirm the multi-omics PCA derived gene-level classification findings, we performed a simple dual dimensionality reduction and differential analysis approach on the LumA vs. LumB comparison with edgeR (Robinson et al., 2010). Briefly, edgeR fits a negative binomial distribution to assess whole-transcriptome gene expression. In this second analysis, the top 44 differentially expressed mRNAs were used for gene-level classification in the same manner as described above. Given edgeR gene-level classification was comparable to PCA gene-level findings (Supplemental Figure S4), we used the features from PC1 to take advantage of the enhanced molecular information content of our multi-omics approach.

Close inspection of the top 44 genes from PC1 used as molecular features for the LumA vs. LumB comparison indicated that *RACGAP1* was the most informative feature, as averaged across nine of the ten classifiers (see Supplemental Spreadsheet 1-S2; RBMs were not included because of the difficulty in assessing feature importance). This finding was further supported via an independent edgeR (Robinson et al., 2010) analysis, which showed *RACGAP1* was the strongest differentially expressed gene (FDR = $2.57 \times 10^{-36}$; logFC = -1.11) of the top 41 mRNA genes. Supplemental Figure S5 presents a rank-ordered heatmap of the averaged state for each of the 44 genes (41 mRNA and 3 methylated genes) across the 100 unique training sets for the LumA vs. LumB comparison. Conversely, *RACGAP1* was ranked only 22 of 44 by PC1 loading. These findings indicate the importance of combined dimensionality reduction/feature learning and classification of high-throughput biological data. From a biological perspective, *RACGAP1* is a putative oncogene, which promotes growth of triple negative/basal-like breast cancers. Experimental depletion of this gene inhibits cancer cell proliferation by the combined effects of cytokinesis failure, *CDKN1A/p21*-mediated *RB1* inhibition, and the onset of senescence (Lawson et al., 2016). Given the significant increased expression of *RACGAP1* in Luminal B tumors, the more aggressive breast cancer subtype, our gene-level classification results also support our previous findings indicating Ising-type models are capable of robustly assigning the greatest weight to the most biologically relevant information in a given model. Figure 4b shows hierarchical clustering of the 44 most informative genes for the LumA vs. LumB breast cancer comparison and indicates significant discrimination between LumA vs. LumB based on these 44 genes.

Finally, we used GOseq analysis (Young et al., 2010) and a PubMed Central (PMC) comprehensive semantic search engine to determine known biological relevance of the top 44 genes in the LumA vs. LumB breast cancer comparison. Our GOseq analysis produced 244 functionally enriched gene ontology (GO) terms (see Supplemental Spreadsheet 1-S3). Of these, Figure 4c presents nine statistically significant (Wallenius approximation; FDR ≤ 0.05) GO terms related to cancer: *metabolic process*; *cell cycle*; *heterocycle metabolic process*; *regulation of the cell cycle*; *glucose 6-phosphate metabolic process*; *DNA integrity checkpoint*; *telomere organization*; and *morphogenesis of a branching epithelium*. We then used a semantic search engine to query full-text records available in PMC database for published relationships between these 44 genes and the query terms, *cancer* and *breast cancer* (see Supplemental Methods). Briefly, we used the *entrez search* function of the *rentrez* R package, which provides an NCBI *EUtils* application programming interface (API) (Winter, 2017), to retrieve results for each of the 44 genes from the PMC database. Search terms were defined by combining each gene symbol with either *cancer* or *breast cancer* fields, along with all related MeSH terms using Boolean operators *AND/OR*.

We found that all but *C12orf73* have been previously indicated in breast cancer (Figure 4d). Of the remaining 43 genes, *PRR15L* and *MAGI2-AS3* are the only genes with no current functional annotation; however, both *PRR15L* and *MAGI2-AS3* associate with a high averaged information ranking for the LumA vs. LumB comparison (see Supplemental Spreadsheet 1-S2). At the time of our semantic search of the PMC database, Hepatocyte Growth Factor (*HGF*) and Retinoblastoma-Associated Protein 1 (*E2F1*) were implicated in the greatest number of published breast cancer papers (6,356 and 5,925, respectively) among all of the 44 genes queried (see Supplemental Spreadsheet 1-S4). *E2F1* yielded higher PC1 loading (4 vs.15) and averaged information (8.6 vs. 33) rankings than *HGF*. *E2F1* is a well-studied transcription factor involved in cell proliferation, differentiation, and apoptosis. It is a member of the E2F protein family, which has been implicated in cell cycle control and regulation of tumor suppressor proteins. Low *E2F1* gene expression is predictive of metastasis-free survival in breast cancer patients (Vuaroqueaux et al., 2007). As with our *RACGAP1* finding, we determined significantly higher differential mRNA expression of *EF2F1* in LumB vs. LumA breast cancers via edgeR analysis (FDR = 2.59 x $10^{-27}$; logFC = -1.34). Taken together our gene-level classification results support known breast cancer etiology.

**Discussion**

We have presented the first successful demonstration of annealing-based Ising models applied to integrated genome-wide, multi-omics human cancer data. We have shown that classification with Ising models is comparable to standard ML strategies on multiple partitions of data of multiple large human cancer datasets. However, it is important to note that the benefit of using quantum annealing cannot be attributed solely to inherent quantum behavior, as simulated annealing and our random control classifier performed similarly if not better than quantum annealing as implemented by a D-Wave device on two of the three fractional training dataset comparisons. By randomly generating bit strings and sorting them by their Ising energy, we achieved classification accuracies nearly equal to standard ML and, in some cases, better than both quantum and simulated annealing. The comparable performance of our random control strategy to D-Wave and SA is due to a distinction between the objective function for the Ising-type approaches, which is an approximation for the negative log-likelihood, and the performance metrics presented (accuracy, balanced accuracy, F1 score, AUC). While we describe this discrepancy in more detail in the Supplemental Methods and in Supplemental Figure S6, we found the overall classification performance of the random classifier a direct indication of the utility of formulating a classification problem as an Ising Hamiltonian. In this current study, the advantage of using an Ising problem became even more apparent by training classifiers on a relatively smaller amount of training data, as we witnessed with the LumA vs. LumB and ERpos vs. ERneg breast cancer comparisons. For example, Field, which is an almost trivial algorithm after formulating the Ising problem, performed extremely well from 95% to 20% of original training data for this breast cancer comparison.

The relative advantage of annealing-based Ising approaches over standard ML approaches when trained with relatively small amounts of data may be attributed to the discrete weights returned for the Ising-type methods. On the one hand, discrete weights rendered with Ising-type methods control for statistical shrinkage better than statistical optimization parameters of standard ML approaches. This generalizability issue has plagued the ML field for decades. On the other hand, binary weights limit the informativeness of the standard classifiers; with larger amounts of training data, the Ising-type methods slightly underperformed standard ML approaches. These findings point to the potential application of a new class of algorithms as simple heuristic models with discrete weights may perform better in situations of limited training data, which is often the case in clinical trials and drug efficacy studies. The relative advantageous trend of enhanced classification performance for Ising-type methods on small amounts of training data are true even when using gene-level features; Supplemental Figure S7 shows balanced accuracies for LumA vs. LumB and ERpos vs. ERneg breast cancer comparisons relative to the top 44 genes from PC1 on incrementally smaller amounts of training data. Interestingly, the gene-level RBM under performs all other Ising-type methods at all but the smallest fraction of training data. Moreover, Supplemental Figure S8 shows statistically enhanced control of overfitting for Ising-type methods, especially at low fractions, on both LumA vs LumB and ERpos vs. ERneg comparisons.

Although RBMs are formulated quite differently from the annealing-based Ising models, in that RBMs are unsupervised and iteratively update model parameters, whereas the annealing-based approaches explored here are supervised with fixed $h$'s and $J$'s, the overall trends in classification performance for all these Ising models is quite similar. When using very small amounts of training data, RBMs also seem to perform much better than the standard ML methods, although their performance is noticeably better on the PCs than on the gene-level features. This may be attributed, in part, to distinct underlying data distributions; different algorithms do better on certain types of data. For example, Field, which performed quite well for the LumA vs LumB comparison on the PC-level data performed poorly on the gene-level data, while RF performed better at the gene-level than the PC-level data. Differences in data type notwithstanding, the reason RBMs are performing well with small amounts of data may also be due to the binary nature of the hidden units. Although there are no "weights" to be learned, as there are with the annealing-based approaches, the hidden units for the RBMs were binarized during training; it is possible that this intentional sacrifice of precision leads to less overfitting with relatively smaller amounts of data.

Inherent to all useful biological classifiers, we showed that all the Ising-type algorithms identified relevant molecular features in each cancer comparison. Like the standard ML approaches, these algorithms determined PC1 as the most informative feature for each dataset, from which we then proceeded to perform gene-level classification. Analysis of feature importance of the trained classifiers on the top 44 genes of PC1 for the LumA vs. LumB comparison determined *RACGAP1*, a putative oncogene in breast cancer, associated with the highest averaged information ranking. This finding was supported via independent differential gene expression analysis, indicating Luminal B tumors, a more aggressive molecular subtype of breast cancer, associated with statistically significant, higher mRNA levels of *RACGAP1* than Luminal A tumors. Moreover, our semantic search of full-text records available in PMC database found that 43 of these top 44 genes have been previously implicated in breast cancer. While our results support previously published findings, it is possible that more sophisticated dimensionality-reduction techniques, such as multi-omics factor analysis (Argelaguet et al., 2018) could be used to provide fresh insights into the mechanisms of disease. The effect of such techniques on the relative performance of both standard and Ising-type machine learning methods is worthy of further study.

While we achieved comparable classification performance on all binomial comparisons assessed in this study, it is important to note our Ising-type approaches did not perform as well as standard ML on a large multiclass, six-class dataset. This observation is most likely related to the relatively larger training data used for this multiclass comparison, as the six-cancer dataset comprised approximately 12 times the amount of data relative to the LumA vs. LumB dataset. As we showed by reducing the amount of training data for the LumA vs. LumB, ERpos vs. ERneg, and the six-cancer multiclass comparisons, the Ising-type approaches performed well with relatively smaller amounts of data but did not statistically improve with incremental increases. Another explanation for the decreased performance of the Ising-type approaches may be related to the fact that the number of approximations used to formulate the classification problem as an Ising Hamiltonian depends on the number of classes (see Supplemental Methods Eq. (15)). The approximation may be valid for binomial comparisons but could break down with multiclass experimental designs. In contrast, the RBMs, which are not formulated based on the same approximations, perform significantly better than the annealing-based Ising models.

Though practical quantum computing architectures are still in development, the demonstration of comparable classification performance of Ising-type approaches with standard ML methods on high-dimensional, multi-omics human cancer datasets is encouraging. Our survey of ML classifiers has uncovered a class of algorithms that perform better than standard methods on limited biomedical data: Ising-type methods with discrete weights. This advantage for small experimental designs is particularly useful in medicine, where large datasets may be prohibitively expensive to obtain, or in the study of rare diseases. As technology improves and new algorithms are introduced, we are cautiously optimistic that these unconventional classification algorithms will afford unique insights and drive the discovery of novel approaches for solving complex biological problems.

**Methods**

*Dataset and dimensionality reduction*

Genomic data from The Cancer Genome Atlas (TCGA) was retrieved, pre-processed, and normalized. An overview of our data pipeline is depicted in Fig. 1. Briefly, we retrieved whole exome sequencing, RNA-Seq, miRNA-Seq, DNA methylation array, and genotyping array data for five human cancer binomial classifications (breast cancer vs. normal, estrogen receptor positive vs. estrogen receptor negative breast cancers, luminal A vs. luminal B breast cancers, kidney renal clear cell vs. papillary cell carcinoma, and lung adenocarcinoma vs. squamous cell carcinoma) as well as a six-cancer multiclass classification, which included breast, colorectal, lung, kidney, brain, and liver cancer types). Data were retrieved from either the Genome Data Commons (GDC) data portal (https://portal.gdc.cancer.gov/ - data release 4.0) or cBioportal (http://www.cbioportal.org/) (Cerami et al., 2012; Gao et al., 2013). All five data types (mRNA, miRNA, CNV, DNA methylation, and somatic tumor variants) were preprocessed independently (see Supplemental Methods) and then concatenated into a single data matrix.

We derived classification performance via 100 random, approximately class-balanced partitions of training (80%) and test/validation (20%) data. Each feature was standardized to zero mean and unit variance (z-score) based on the training data. The same training mean and standard deviation was then applied to the corresponding test data. Furthermore, given that the data comprised more than 79,000 molecular features, dimensionality reduction was conducted in order to make comparisons with existing quantum hardware. As such, we performed PCA on each random, balanced partition of the training data, retaining the top 44 principal components for the binomial datasets and 13 principal components for the six-cancer dataset. The test data was then projected onto the PCs defined by the corresponding training data. The number of principal components was chosen based on the largest number of features that could be accommodated on existing quantum annealing hardware (see the Section below on formulating the classification problem as an Ising model). Hyper-parameter was selected using cross-validation on the training data (see Supplemental Information for more information about which hyper-parameters were chosen).

### *Quantum annealing*

We explored the use of quantum annealing with processors made by D-Wave Systems Inc (Johnson et al., 2011; Lanting et al., 2014) (see the Supplemental Methods for a brief review of quantum annealing). Results for the binomial datasets were obtained by running the D-Wave 2X (DW2X) processor installed at the Information Sciences Institute (ISI) of the University of Southern California, and results for the six-cancer dataset were run on the DW2000Q located in Burnaby, Canada. The problem Hamiltonians that are used for D-Wave (DW) can be described as Ising spin models with tunable parameters (Johnson et al., 2011). The Ising model assumes a graph $\mathcal{G} = (V, E)$ composed of a set of vertices, $V$, and edges, $E$. Each of the spins is a binary variable located at a unique vertex. The $N$ spins are represented by superconducting flux qubits, and $G$ is the so-called Chimera graph (see Supplemental Figure S9). For the DW2X, $N = 1098$ and for the DW2000Q, $N = 2038$. The problem (or Ising) Hamiltonian for this system can be written as

$$H_P = \sum_{i \in V} h_i \, \sigma_i^z + \sum_{(i,j) \in E} J_{i,j} \, \sigma_i^z \sigma_j^z$$

(1)

where the local fields $\{h_i\}$ and couplings $\{J_{ij}\}$ define a problem instance and are programmable on the DW2X to within a few percent Gaussian distributed error. The $\{\sigma_i^z\}$ represent both binary variables taking on values $\pm 1$, and the Pauli $z$-matrices. Given a spin configuration $\{\sigma_i^z\}$, $H_P$ is the total energy of the system. Problems submitted to DW are automatically scaled so that all $h_i$ and $J_{ij}$ values lie between $-1$ and 1. The initial Hamiltonian is $H_B = \sum \sigma_i^x$ a transverse magnetic field where is the Pauli $x$-matrix acting on qubit $i$. During an anneal, the magnitude of $H_B$ is gradually reduced to zero, while the magnitude of $H_P$ is slowly increased from zero. After each anneal DW returns a set of spin values $\{\sigma_i^z = \pm 1\}$ that attempts to minimize the energy given by Eq. (1) (a lower energy indicates better optimization). Note, however, that for our purposes we are not strictly using DW as an optimizer. In the Supplemental Methods, we describe our procedure to make use of the fact that higher-energy solutions may still contain some meaningful information and use them to improve performance.

For the results in the main text, we set the annealing time at $5\mu s$ and repeated the anneal 1000 times, which returns 1000 spin configurations. We selected the 20 spin configurations with the lowest Ising energy and ran some quick classical post-processing to average the lowest Ising energy spin configurations if they improved the objective function on the training data. See the Supplemental Methods for a more detailed description of other hyper-parameters and Supplemental Figures S10-S12 for the effect of using a larger number of spin configurations.

### Simulating Annealing

Similar to quantum annealing, simulated annealing (SA) accepts problems formulated as an Ising problem, as defined in Eq. (1) and returns binary variables. For this work we used the implementation of Isakov *et al*. (Isakov et al., 2015). There are several important parameters that affect SA's overall performance: the number of sweeps, the type of schedule (linear or exponential in the inverse temperature), and the initial and final temperatures. For our purposes, we fixed the number of sweeps (which is analogous to the annealing time of quantum annealing) to 1000 and selected a linear schedule with an inverse initial temperature of 0.01. We treated the final inverse temperature as a tunable hyper-parameter with values in the set 0.03,0.1,0.3,1,3 and repeated the anneal 1000 times. Results in the main text are given for the final inverse temperature that yielded the best performance during cross-validation. We used the same classical post-processing procedure that was used with D-Wave to combine 20 spin configurations with the lowest energy, not just the one that returned the lowest Ising energy.

### Field

As another approach to explore the usefulness of the formulating the classification task as an Ising problem, and to check the role played by the couplings (the $J$'s) we implemented a very simple algorithm that only takes into account the values of the local fields (the $h$'s) in Eq. (1). Once $h$ has been determined based on the training data, we choose the weights to be the opposite sign of the fields; i.e., $\sigma_i^{\text{field}} = -h_i$. This amounts to a (trivial) analytical solution of the optimization of Eq. (1) without any $J$'s.

### Random

As a sanity check, we generated random solutions to Eq. (1). For each spin we picked a random number uniformly distributed in the interval $[0,1)$. Values below 0.5 were set to -1 and those above 0.5 were set to 1, thereby generating spin configurations the same as those returned by DW and SA. We then sorted the spin configurations according to their Ising energy, given by Eq. (1). As with DW and SA, we generated 1000 such random spin configurations and used the same classical post-processing procedure to combine the 20 spin configurations with the lowest energy to a final set of weights.

### RBM

Restricted Boltzmann machines (RBMs) are a class of bipartite unsupervised energy-based models that consist of a visible layer (the data one would like to model) and a hidden layer. Introducing a hidden layer allows one to model more complicated probability distributions. An RBM defines a probability distribution through an energy function as: $Pr(x) = exp[-H(x)]/Z$ , where $Z = \sum_x[-H(x)]$ The energy function for RBMs can be written as an Ising Hamiltonian, though the goal of learning is to update the $h$'s and $J$'s such that the probability distribution most closely resembles the input data distribution; in contrast, the $h$'s and $J$'s for the models above are fixed, given the training data.

In order to use RBMs for classification, we used the approach described in (Larochelle et al., 2012), where class labels were added to the visible layer and the RBMs learned to model a distribution over both the "data" units and the "label" units, to adapt the R 'deepnet' package. After training, the probability of each class was calculated by setting the corresponding label unit to 1 and all other label units to 0. The RBMs were trained with contrastive divergence using a $k = 1$, and a batch size of 32.

### *Formulating a multiclass classification problem on a quantum annealer*
We show how to arrive at a simple Ising formulation to model a multiclass classification problem with $K$ unique class labels. Assume we have a dataset of $N$ training examples: $D = (x_i, y_i)_{i=1}^{N}$ where $x_i$ is the $i$th data vector of $M$ features and $y_i$ is the corresponding class of the $i$th data vector (i.e., $y_i$ can take one of the $K$ class labels). A simple way to arrive at probabilities for a multiclass classification problem is to use the softmax function. We can define the probability of each class as

$$Pr(y_i = k) = \frac{exp\ w_k^\intercal x_i}{\sum_{k=1}^{K} exp\ w_k^\intercal x_i}$$

(2)

where $w_k$ are the weights corresponding to the $k$th class that we would like to learn (in other words, we define a set of weights for each class). However, since we are generating a probability of each class, we can reduce the set of weights we have to train from $K$ to $K - 1$ and define the first $K - 1$ probabilities as:

$$Pr(y_i = k) = \frac{exp\ w_k^\intercal x_i}{1 + \sum_{k=1}^{K-1} exp\ w_k^\intercal x_i}$$

(3)

with the probability of the $K$th class as:

$$Pr(y_i = K) = \frac{1}{1 + \sum_{k=1}^{K-1} exp\ w_k^\intercal x_i}$$

(4)

The goal of training is to find the weights that maximize the probability given the classes in the dataset, or equivalently to minimize the negative log-likelihood. Once the weights are found, inference is straightforward; probabilities for each class are generated and we assign the predicted label based on the class with the highest predicted probability. We can express the negative log-likelihood as follows:

$$\mathcal{L} = -\log \prod_i Pr(y_i)$$

(5)

$$= -\sum_i \log Pr(y_i)$$

(6)

where the probability selected corresponds to the actual class of the label. If the actual class has the highest predicted probability for all data samples, the negative log-likelihood will be minimized. In other words, the farther away from 1 the predicted probability of the real class is, the greater the contribution to the negative log-likelihood; if the algorithm were able to correctly assign a class to each training example with probability 1, the negative log-likelihood would be 0.

Taking a second-order Taylor approximation around the argument of the exponential equal to 0, we eventually arrive at the following expression for the negative log-likelihood (see Supplemental Methods for a more complete derivation and additional technical concerns):

$$\mathcal{L} \approx \sum_{k=1}^{K-1} w_k^\mathsf{T}(b_k + h) + \sum_{k=1}^{K-1} w_k^\mathsf{T} J' w_k - \sum_{k=1}^{K-1} \sum_{j \neq k} w_j^\mathsf{T} J'' w_k \,,$$

(7)

where

$$b_k = \sum_{i:y_i=k} -x_i \,, h = \frac{1}{K} \sum_i x_i \,,$$

(8)

$$J' = \frac{K-1}{2K^2} \sum_i x_i x_i^\mathsf{T} \,, J'' = \frac{1}{2K^2} \sum_i x_i x_i^\mathsf{T} \,.$$

(9)

In general, this formulation requires arbitrary inter-weight couplings (i.e., $J''$-- couplings between $w_k$ and $w_j$ where $k$ and $j$ represent the vector of weights for classes $k$ and $j$) and intra-weight couplings ($J'$-- couplings between $w_{k,n}$ and $w_{k,m}$, where $n$ and $m$ are the indices of the weights assigned to the $n$th and $m$th features for the vector of weights for the $k$th class). This imposes constraints on the number of classes and number of features that can be run on a particular hardware graph. For a dataset with $M$ features and $K$ classes, this approach requires $M \times (K - 1)$ logical variables, and in general there must be . Due to restrictions on the number of qubits and the limited graph connectivity (see Figure S9), an *embedding*, by which edges in the graph are contracted to give a graph with fewer vertices but a higher degree (See Supplemental Information for more details). For the D-Wave 2000Q, the largest complete graph that can be embedded (Choi, 2008) consists of 66 logical variables; i.e., $M \times (K - 1)$ must be at most 66. For our purposes, we chose $K = 6$ cancer types, which limits the number of features we can use to 13. The largest complete graph that can be embedded onto the DW2X processor at ISI consists of 45 logical variables, so for the binomial datasets we chose a total of 44 features.

## Data Availability
The data that supports the findings of this study is available via SharePoint link in GitHub https://github.com/wuxi-nextcode/quantumML.

## Acknowledgements
We would like to thank D-Wave for access to their 2000Q processor in Burnaby, Canada.

## Author Contributions
T.W.C. and D.A.L. conceived the project. R.Y.L., J.R.G., D.A.L., and T.W.C., conceived the experimental design and wrote the manuscript. R.Y.L., S.G., S.R.B., N.C., and O.E.G. processed, scaled, and analyzed all data.

## Competing Interests
O.E.G, S.G., J.R.G., N.C., S.R.B., and T.W.C. were employed by Genuity science during the research project. R.Y.L. was the recipient of a research grant from Genuity Science during the research project. The work of D.A.L. is based upon work (partially) supported by the Office of the Director of National Intelligence (ODNI), Intelligence Advanced Research Projects Activity (IARPA), via the U.S. Army Research Office contract. The views and conclusions contained herein are those of the authors and should not be interpreted as necessarily representing the official policies or endorsements, either expressed or implied, of the ODNI, IARPA, or the U.S. Government. The U.S. Government is authorized to reproduce and distribute reprints for Governmental purposes notwithstanding any copyright annotation thereon. The authors declare no other competing interests.

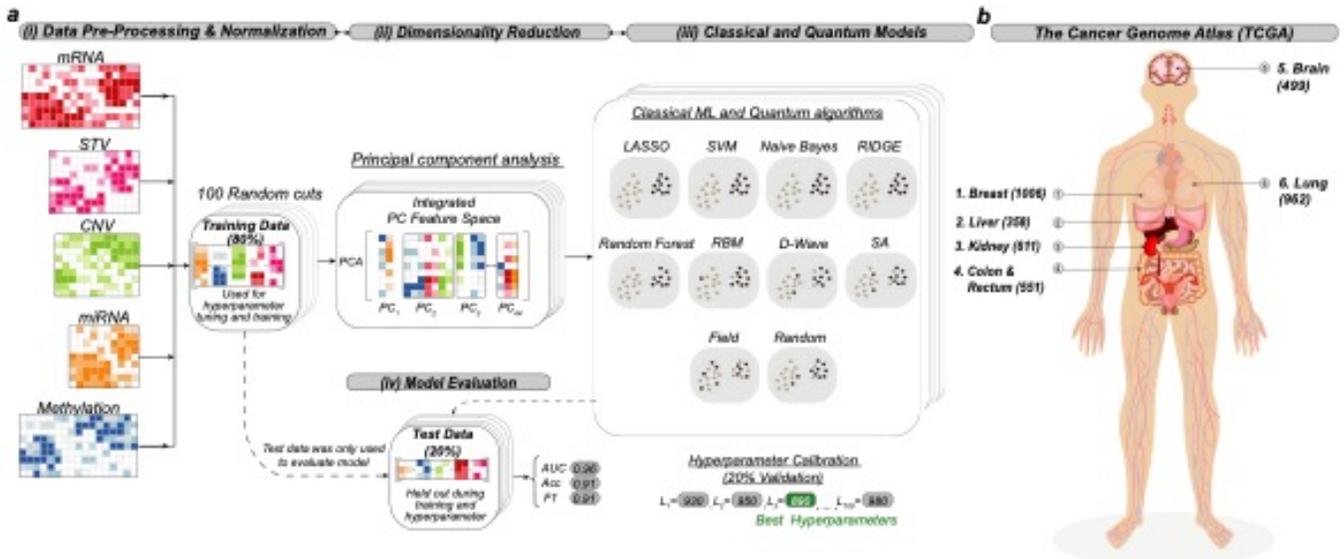

**Figure 1**: (a) Overview of classification strategy. (i) Whole exome sequencing, RNA-Seq, miRNA-Seq, DNA methylation array, and genotyping array data (for CNVs) were retrieved from The Cancer Genome Atlas (TCGA) for human cancer type and molecular subtype classification. Data were concatenated and transformed into a single scaled 'omics' data matrix. The matrix was then repeatedly split into 100 unique training and independent test sets representing 80% and 20% of total data, respectively. After splitting the data, each training split was scaled to have zero mean and unit standard deviation. The same scaling was then applied to the corresponding test split. (ii) Principal Component Analysis (PCA) was performed separately on each individual training set, and a subsequent matched test set was projected using training set specific PCA loadings. (iii) Several standard classical machine learning algorithms were compared to quantum annealing and several classical algorithms that have the same objective function as quantum annealing. The standard classical machine learning methods assessed included: Least Absolute Shrinkage and Selection Operator (LASSO), Ridge regression (Ridge), Random Forest (RF), Naïve Bayes (NB), and support vector machine (SVM). Quantum annealing (D-Wave) was performed on D-Wave hardware by formulating the classification problem as an Ising problem (see Methods). These classical Ising-type approaches include: simulated annealing (SA), candidate solutions randomly generated and sorted according to the Ising energy (Random), and an approach that considers only local fields of the Ising problem (Field). Hyper-parameters were tuned on the train data using a 10-fold cross-validation (see Supplemental Methods for a description of the ranges of hyper-parameters used). (iv) After training, classification performance was validated with each corresponding test set (unseen during the tuning of hyper-parameters and the training) for a variety of statistical metrics, including balanced accuracy, area under the ROC curve (AUC), and F1 score. Classification performance metrics were averaged for the 100 test sets for each model to provide statistics on the mean performance. (b) Presents the six human cancer types used for the multiclass classification models. Patient sample sizes are indicated in parentheses.

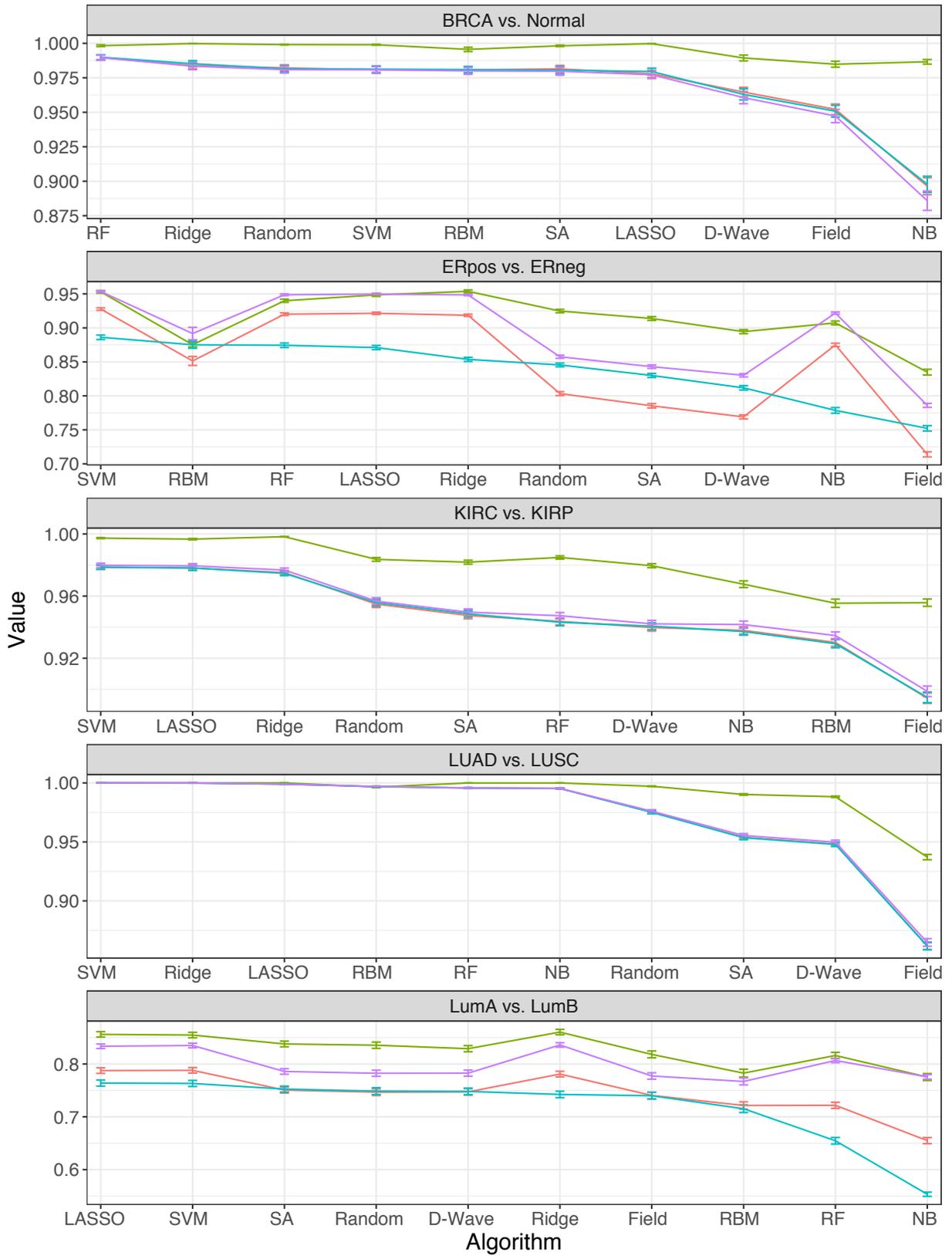

**Figure 2**: Comparison of classification algorithms for five TCGA cancer datasets. Human cancer datasets assessed: Kidney Renal Clear Cell Carcinoma (KIRC) vs. Kidney Renal Papillary Cell Carcinoma (KIRP); Lung Adenocarcinoma (LUAD) vs. Lung Squamous Cell Carcinoma (LUSC); Breast Invasive Carcinoma (BRCA) vs. matched normal tissue (normal); estrogen receptor positive (ERpos) vs. estrogen receptor negative (ERneg) breast cancers; and luminal A (LumA) vs. luminal B (LumB) breast cancers. To address class imbalance for each comparison, algorithm performance is ranked by mean balanced accuracy on the x-axis. By and large the other metrics indicate the same performance ranking. Classification performance metrics were averaged for the 100 unique training and test sets for each model (see Methods). Performance metrics: Accuracy (red), AUC (green), Balanced Accuracy (blue), F1 score (purple). Data are presented as mean ± SEM.

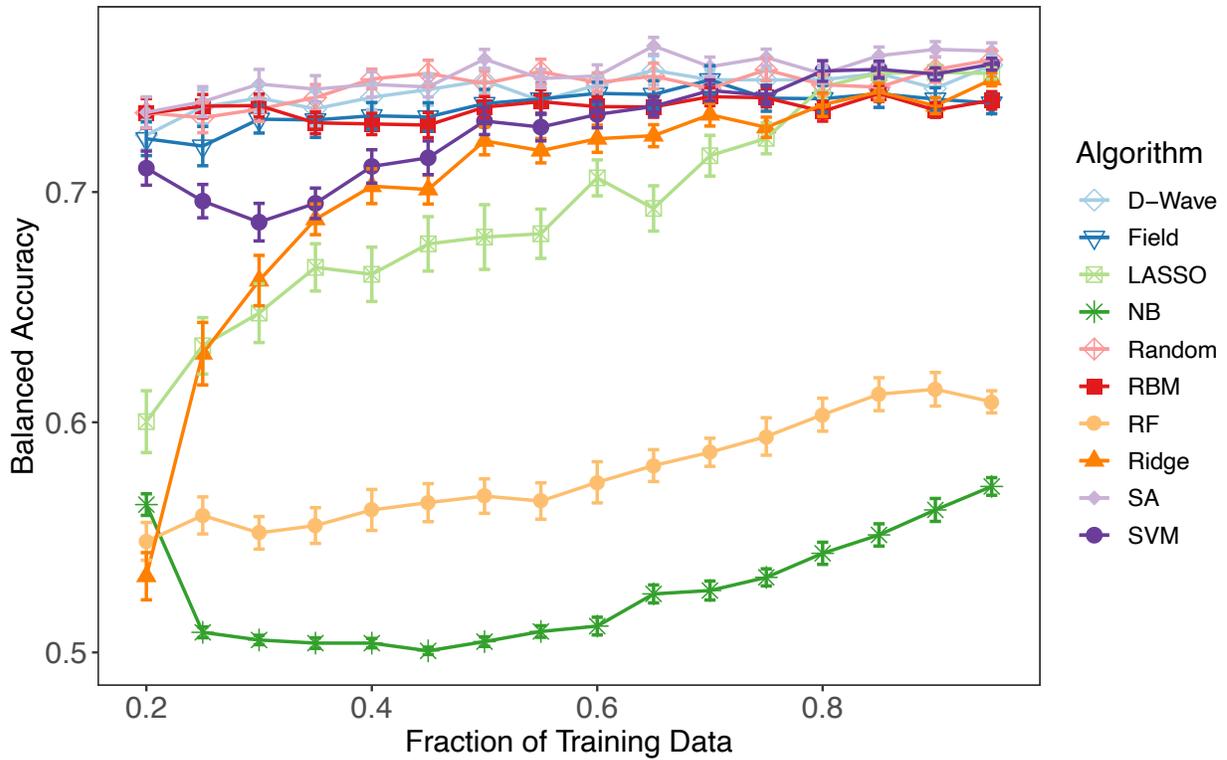

**Figure 3**: Test set balanced accuracy for LumA vs. LumB binomial classification with incremental decreases from 95% to 20% of original training set. Algorithms evaluated are indicated in the legend. Averaged balanced accuracies were calculated for 50 independent training sets at each designated fraction of original training data. Data are presented as mean ± SEM.

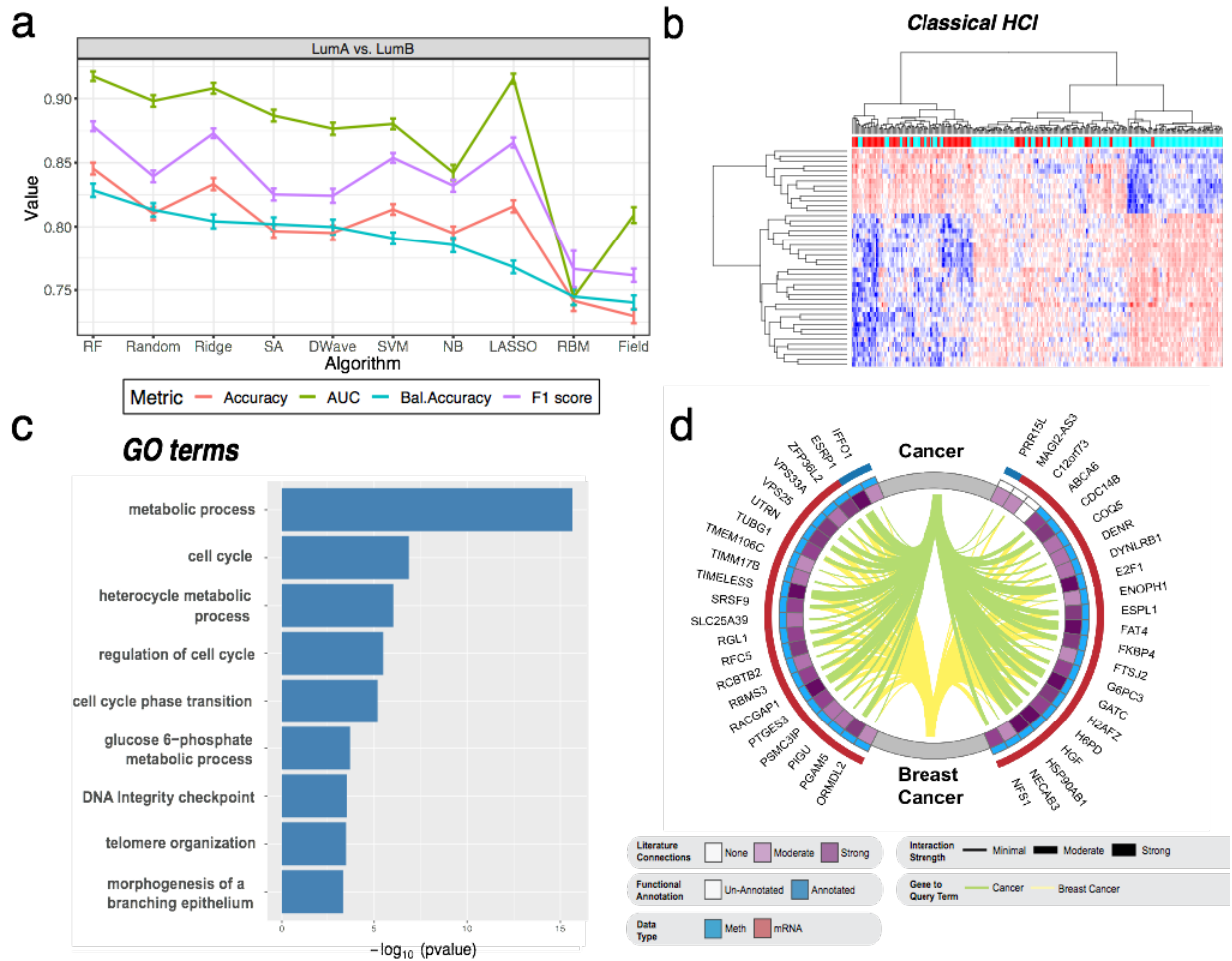

**Figure 4**: Classification, hierarchical clustering, functional enrichment, and natural language processing of the top 44 genes of PC1 for Luminal A vs. Luminal B binomial comparison. (a) Gene-level classification of Luminal A vs. Luminal B human breast cancers based on the top 44 genes of PC1. Data presented as mean ± SEM. (b) classical HCL algorithm (see Methods). Note: genes are presented in rows and samples in columns. (c) GOseq functional enrichment analysis of top 44 genes for PC1 shows enriched GO terms ordered by p-values. (d) Circos plot representing semantic search of full-text articles within the PubMed Central Database identifying published associations of the top 44 genes for PC1 to the query terms, *Cancer* and *Breast Cancer*. The red and blue outer bands represent 'mRNA' and 'methylation' datatypes, respectively. The inner blue band represents genes with known functional annotation. The intensity of the inner purple-colored ring indicates the total number of publications in cancer and breast cancer of the top 44 genes of PC1. This banned colored with 6 bins where white is the lowest and dark purple the highest number of publications at the time of analysis. The thickness and color of the circus plot ribbons indicate number of published gene-to-query term associations: green represents cancer and yellow designates breast cancer.

| Dataset | LASSO | Ridge | SVM | RF | NB | DW | SA | Rand | Field | RBM |
|---|---|---|---|---|---|---|---|---|---|---|
| BRCA vs Normal | 0.982 ± 0.002 | .984 ± 0.002 | 0.981 ± 0.002 | 0.990 ± 0.002 | 0.897 ± 0.006 | 0.974 ± 0.003 | 0.981 ± 0.003 | 0.982 ± 0.002 | 0.951 ± 0.004 | 0.981 ± 0.002 |
| ER pos vs ER neg | 0.871 ± 0.003 | 0.854 ± 0.003 | 0.886 ± 0.003 | 0.874 ± 0.003 | 0.779 ± 0.004 | 0.812 ± 0.003 | 0.830 ± 0.003 | 0.845 ± 0.003 | 0.752 ± 0.004 | 0.875 ± 0.005 |
| KIRC vs KIRP | 0.978 ± 0.002 | 0.975 ± 0.002 | 0.979 ± 0.001 | 0.944 ± 0.002 | 0.937 ± 0.002 | 0.947 ± 0.002 | 0.949 ± 0.002 | 0.956 ± 0.002 | 0.895 ± 0.003 | 0.929 ± 0.002 |
| LUAD vs LUSC | 0.9988 ± 0.0002 | 0.9999 ± 0.0001 | 1.0000 ± 0.0000 | 0.9957 ± 0.0004 | 0.9953 ± 0.0004 | 0.9478 ± 0.0017 | 0.9536 ± 0.0017 | 0.9751 ± 0.0013 | 0.8616 ± 0.0031 | 0.997 ± 0.0004 |
| Lum A vs Lum B | 0.764 ± 0.006 | 0.742 ± 0.006 | 0.763 ± 0.006 | 0.655 ± 0.006 | 0.553 ± 0.004 | 0.748 ± 0.006 | 0.752 ± 0.006 | 0.749 ± 0.006 | 0.740 ± 0.006 | 0.715 ± 0.007 |
| 6 cancer | 0.9896 ± 0.0002 | 0.9845 ± 0.0003 | 0.9891 ± 0.0002 | 0.9863 ± 0.0003 | 0.9735 ± 0.0004 | 0.9122 ± 0.0012 | 0.9170 ± 0.0010 | 0.9083 ± 0.0015 | 0.8876 ± 0.0005 | 0.9534 ± 0.0013 |

**Table 1:** Balanced accuracies for five binomial comparisons and the one six-class cancer dataset used in this study. For the six cancer dataset. Data reported as mean ± SEM

| Fraction | LASSO | Ridge | SVM | RF | NB | DW | SA | Rand | Field | RBM |
|---|---|---|---|---|---|---|---|---|---|---|
| 0.25 | 0.633 ± 0.012 | 0.630 ± 0.013 | 0.696 ± 0.007 | 0.560 ± 0.008 | 0.508 ± 0.002 | 0.737 ± 0.007 | 0.739 ± 0.007 | 0.732 ± 0.007 | 0.720 ± 0.009 | 0.737 ± 0.005 |
| 0.55 | 0.682 ± 0.011 | 0.718 ± 0.005 | 0.728 ± 0.006 | 0.566 ± 0.008 | 0.509 ± 0.003 | 0.740 ± 0.006 | 0.749 ± 0.004 | 0.752 ± 0.005 | 0.740 ± 0.007 | 0.739 ± 0.005 |
| 0.95 | 0.752 ± 0.002 | 0.749 ± 0.003 | 0.756 ± 0.003 | 0.609 ± 0.005 | 0.572 ± 0.004 | 0.755 ± 0.004 | 0.761 ± 0.004 | 0.757 ± 0.004 | 0.739 ± 0.005 | 0.740 ± 0.004 |

**Table 2:** Balanced accuracies for when incrementally decreasing the amount of training for the Luminal A versus Luminal B comparisons. The fraction represents the amount of training data used, and results are reported as mean ± SEM.